\documentclass[%
reprint,
superscriptaddress,
nofootinbib,
amsmath,amssymb,
prc,
floatfix,
]{revtex4-2}

\usepackage{graphicx}
\usepackage{dcolumn}
\usepackage{bm}
\usepackage{color}
\usepackage{here}
\usepackage{hyperref}
\usepackage{nicefrac}
\usepackage{longtable}
\usepackage{lipsum}
\usepackage{ulem}
\usepackage{multirow}
\usepackage[T1]{fontenc}

\newcommand{\nuc}[2]{\hbox{$^{#1}$#2}}

\newcommand{\twoninesixsix}{$2966(3)$}
\newcommand{\threefivezerotwo}{$3502(3)$}
\newcommand{\fourzeroonethree}{$4016(3)$}
\newcommand{\fournineninetwo}{$4991(7)$}
\newcommand{\fivefivefivethree}{$5552(4)$}

\newcommand{\fivethreesix}{$537(4)$}
\newcommand{\onezerofournine}{$1050(3)$}
\newcommand{\onethreesevenzero}{$1369(5)$}
\newcommand{\onefiveeightseven}{$1586(3)$}
\newcommand{\onesevenzerotwo}{$1705(5)$}
\newcommand{\oneseventhreezero}{$1730(3)$}
\newcommand{\oneseveneightfour}{$1782(5)$}
\newcommand{\twozerotwofive}{$2027(6)$}
\newcommand{\twozerofiveone}{$2050(6)$}
\newcommand{\twoonetwoone}{$2120(3)$}
\newcommand{\twoonesevenzero}{$2162(7)$}
\newcommand{\twotwoeightfive}{$2284(3)$}
\newcommand{\twofiveeightseven}{$2586(5)$}
\newcommand{\twosevenfivezero}{$2755(7)$}
\newcommand{\threetwothreenine}{$3249(8)$}
\newcommand{\threesixonetwo}{$3609(11)$}

\begin{document}


\title{In-beam $\gamma$-ray spectroscopy of negative-parity states of \nuc{37}{K} populated in dissipative reactions}

%
\author{T.~Beck}
\email{beck@frib.msu.edu}
\affiliation{Facility for Rare Isotope Beams, Michigan State University, East Lansing, Michigan 48824, USA}
\author{A.~Gade}
\affiliation{Facility for Rare Isotope Beams, Michigan State University, East Lansing, Michigan 48824, USA}
\affiliation{Department of Physics and Astronomy, Michigan State University, East Lansing, Michigan 48824 USA}
\author{B.A.~Brown}
\affiliation{Facility for Rare Isotope Beams, Michigan State University, East Lansing, Michigan 48824, USA}
\affiliation{Department of Physics and Astronomy, Michigan State University, East Lansing, Michigan 48824 USA}
\author{D.~Weisshaar}
\affiliation{Facility for Rare Isotope Beams, Michigan State University, East Lansing, Michigan 48824, USA}
\author{D.~Bazin}
\affiliation{Facility for Rare Isotope Beams, Michigan State University, East Lansing, Michigan 48824, USA}
\affiliation{Department of Physics and Astronomy, Michigan State University, East Lansing, Michigan 48824 USA}
\author{K.W.~Brown}
\affiliation{Facility for Rare Isotope Beams, Michigan State University, East Lansing, Michigan 48824, USA}
\affiliation{Department of Chemistry, Michigan State University, East Lansing, Michigan 48824 USA}
\author{R.J.~Charity}
\affiliation{Department of Chemistry, Washington University, St. Louis, Missouri 63130 USA}
\author{P.J.~Farris}
\affiliation{Facility for Rare Isotope Beams, Michigan State University, East Lansing, Michigan 48824, USA}
\affiliation{Department of Physics and Astronomy, Michigan State University, East Lansing, Michigan 48824 USA}
\author{S.A.~Gillespie}
\affiliation{Facility for Rare Isotope Beams, Michigan State University, East Lansing, Michigan 48824, USA}
\author{A.M.~Hill}
\affiliation{Facility for Rare Isotope Beams, Michigan State University, East Lansing, Michigan 48824, USA}
\affiliation{Department of Physics and Astronomy, Michigan State University, East Lansing, Michigan 48824 USA}
\author{J.~Li}
\affiliation{Facility for Rare Isotope Beams, Michigan State University, East Lansing, Michigan 48824, USA}
\author{B.~Longfellow}
\altaffiliation[Present address: ]{Lawrence Livermore National Laboratory, Livermore, California 94550, USA}
\affiliation{Facility for Rare Isotope Beams, Michigan State University, East Lansing, Michigan 48824, USA}
\affiliation{Department of Physics and Astronomy, Michigan State University, East Lansing, Michigan 48824 USA}
\author{W.~Reviol}
\affiliation{Physics Division, Argonne National Laboratory, Argonne, Illinois 60439 USA}
\author{D.~Rhodes}
\altaffiliation[Present address: ]{Lawrence Livermore National Laboratory, Livermore, California 94550, USA}
\affiliation{Facility for Rare Isotope Beams, Michigan State University, East Lansing, Michigan 48824, USA}
\affiliation{Department of Physics and Astronomy, Michigan State University, East Lansing, Michigan 48824 USA}

\date{\today}

\begin{abstract}

In-beam $\gamma$-ray spectroscopy was used to study excited states of the neutron-deficient nucleus \nuc{37}{K}
populated in fast-beam inelastic-scattering and proton-removal reactions at high-momentum loss.
New $\gamma$-ray transitions and $\gamma\gamma$ coincidence relationships
were established using the $\gamma$-ray tracking array GRETINA.
The extension of the level scheme up to the first $(13/2^-)$ state highlights the potential of this recently demonstrated  
population pathway for studies of isospin symmetry involving mirror-energy differences.
The nature of the newly identified states is discussed in comparison to shell-model calculations with the FSU cross-shell effective interaction. 
The calculated occupation numbers of individual orbitals are shown to offer a consistent explanation 
of the measured mirror-energy differences between \nuc{37}{K} and \nuc{37}{Ar}.

\end{abstract}

\pacs{Valid PACS appear here}
\maketitle

\section{Introduction}
The approximate charge independence of the two-body $NN$ interaction \cite{Mac01b} underpins the concept of isospin \cite{Hei32a,Wig37a} 
and leads to nearly identical level schemes of mirror nuclei, i.e. pairs of nuclei with exchanged numbers of protons $Z$ and neutrons $N$. 
If the charge independence of the strong interaction was perfect, the excitation level schemes of mirror nuclei would be identical. 
Mirror energy differences (MEDs) across a pair of mirror nuclei now arise from Coulomb effects and weak isospin non-conserving interactions \cite{Zuk02a}. 
Examining the drivers behind the magnitude of MEDs has become a rich field of exploration over the past decades \cite{War06a,Ben07a,Ben15a,Ben22a}. 
Recent examples of studies of mirror symmetry in excitation spectra tackled the heaviest pair (\nuc{79}{Zn}/\nuc{79}{Y}) yet \cite{Lle20a}, 
one where both mirror partners are unbound (\nuc{8}{C}/\nuc{8}{He}) \cite{Koy24a}, 
and the case of a  ``colossal'' MED for the $A=36$ pair \nuc{36}{Ca}/\nuc{36}{S} \cite{Val18a,Lal22a}.  

The present work is concerned with the high-spin states of the mirror pair \nuc{37}{K}/\nuc{37}{Ar}, 
in the proximity of the extreme case of $A=36$ referenced above. 
The $A=36,37$ pairs are challenging in two ways: 
(i) the neutron deficient partner of the pair is located near the proton dripline, 
making excited-state spectroscopy challenging from the experimental point of view and 
(ii) with respect to the theoretical interpretation, they are located in the upper $sd$ shell 
where intruder states from the $fp$ shell complicate a shell-model interpretation. 

It was demonstrated recently that reactions induced by fast beams of rare isotopes 
populate complex-structure higher-spin states in the reaction products that underwent a large momentum loss \cite{Gad22a,Gad22b}. 
Such reactions are exploited here for the population of final states in \nuc{37}{K} in two reaction channels, 
one-neutron removal from \nuc{38}{Ca} and \nuc{37}{K} inelastic scattering off \nuc{9}{Be}, both at high momentum loss. 
This new work on high-spin states of \nuc{37}{K} adds to the sparse body of data in the upper $sd$ shell 
where large MEDs have been reported for the $J^{\pi}=13/2^-$ states of the $A=35$ and $39$ mirror pairs 
\nuc{35}{Ar}/\nuc{35}{Cl}~\cite{Ekm04a,Ved07a} and \nuc{39}{Ca}/ \nuc{39}{K}~\cite{And99a}, respectively.  
In the lower $sd$ shell, extensive studies of the mirror pairs \nuc{23}{Mg}/\nuc{23}{Na}~\cite{Jen13a,Bos18a} 
and \nuc{31}{S}/ \nuc{31}{P}~\cite{Jen05a,Tes21a} are available.

In the literature to date, excited states of \nuc{37}{K} for the purpose of spectroscopy were mainly populated in proton-adding 
and nucleon-removing transfer reactions from stable \nuc{36}{Ar} and \nuc{40}{Ca}, respectively.
Together with results from $\beta$ decay of \nuc{37}{Ca}~\cite{Gar91b,Kal97a}, they yielded 
information on excited states with angular-momentum quantum numbers up to $J=7/2$.
Above the low proton-decay threshold of $S_p(\nuc{37}{K})=1857.63(9)$\,keV~\cite{Wan21a},
their depopulation was found to proceed via $\gamma$-ray and proton emission~\cite{Esc88a,Mag95a}.
The data from the $\gamma$-ray spectroscopy presented here is complementary 
since it provides first information on higher-spin, negative-parity states hitherto unknown for \nuc{37}{K}. 
For the mirror partner \nuc{37}{Ar}, ample spectroscopic information on excited states is available~\cite{NDS37}, 
including a recent expansion of the level scheme by states up to about $10.6$\,MeV  and angular-momentum quantum number $J=21/2$
populated in the \mbox{\nuc{27}{Al}(\nuc{12}{C}, $np$)\nuc{37}{Ar}} fusion-evaporation reaction~\cite{Das20a}.


\section{Experiment and Results}
\label{sec:res}

\begin{figure}[t]
\centering
\includegraphics[trim=0 0 0 0,width=1.025\linewidth,clip]{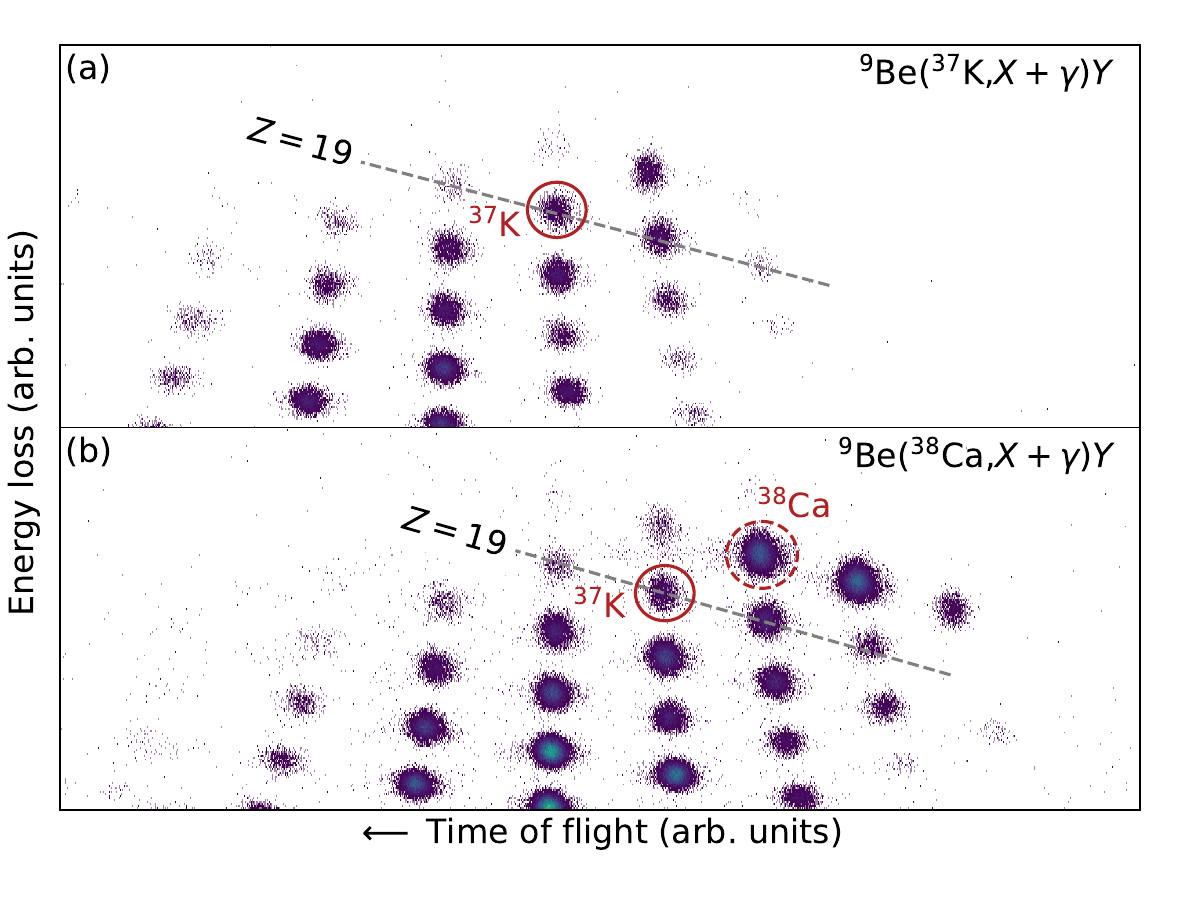}
\caption[]{Event-by-event particle identification plots of projectile-like residues produced 
in the (a) \nuc{9}{Be}(\nuc{37}{K},$X+\gamma$)$Y$ and (b) \nuc{9}{Be}(\nuc{38}{Ca},$X+\gamma$)$Y$ reactions. 
The data displayed in the figure has a particle-$\gamma$ coincidence condition applied. 
}
\label{fig:pid}
\end{figure}

The experimental setup used in this measurement was the same as in Refs.~\cite{Gad20a,Gad22b,Gad22a,Bec23a},
which report results from additional reaction channels, and is briefly introduced in the following.
By fragmenting a stable \nuc{40}{Ca} beam on a $799$-mg/cm$^2$ \nuc{9}{Be} production target, 
accelerated to $140$\,MeV/u by the Coupled Cyclotron Facility of the National Superconducting Cyclotron Laboratory~\cite{Gad16a}, 
the secondary beam containing  \nuc{37}{K} and \nuc{38}{Ca} projectiles was produced. 
Using a $300$\,mg/cm$^2$ aluminum wedge degrader, 
they were separated from other reaction residues in the A1900 fragment separator~\cite{Mor03a}, 
with the momentum acceptance limited to $\Delta p/p=0.25\%$.
The resulting secondary beam comprised \nuc{37}{K} and \nuc{38}{Ca} with shares of approximately $11$ and $85\%$, respectively.
The secondary beam was impinged on a $188$\,mg/cm$^2$ thick \nuc{9}{Be} foil 
located at the target position of the S800 magnetic spectrograph~\cite{Baz03a}, 
with midtarget projectile energies of $58.0$ and $60.9$\,MeV/u, respectively, corresponding to velocities $v/c\approx0.337$ and $0.345$.
The results of the event-by-event particle identification, which show a clear separation of different reaction products, are displayed in Fig.~\ref{fig:pid}.
In the entrance channel, the incoming species was identified via the time-of-flight difference taken between two plastic scintillators 
located at the end of the A1900 and the object position of the S800 analysis beam line.
The outgoing reaction residues were identified from their energy loss in the spectrograph's ionization chamber 
and the time of flight measured between the object position of the analysis beam line and the S800 focal plane~\cite{Yur99a}.

\begin{figure}[tb]
\centering
\includegraphics[trim=0 0 0 0,width=1.025\linewidth,clip]{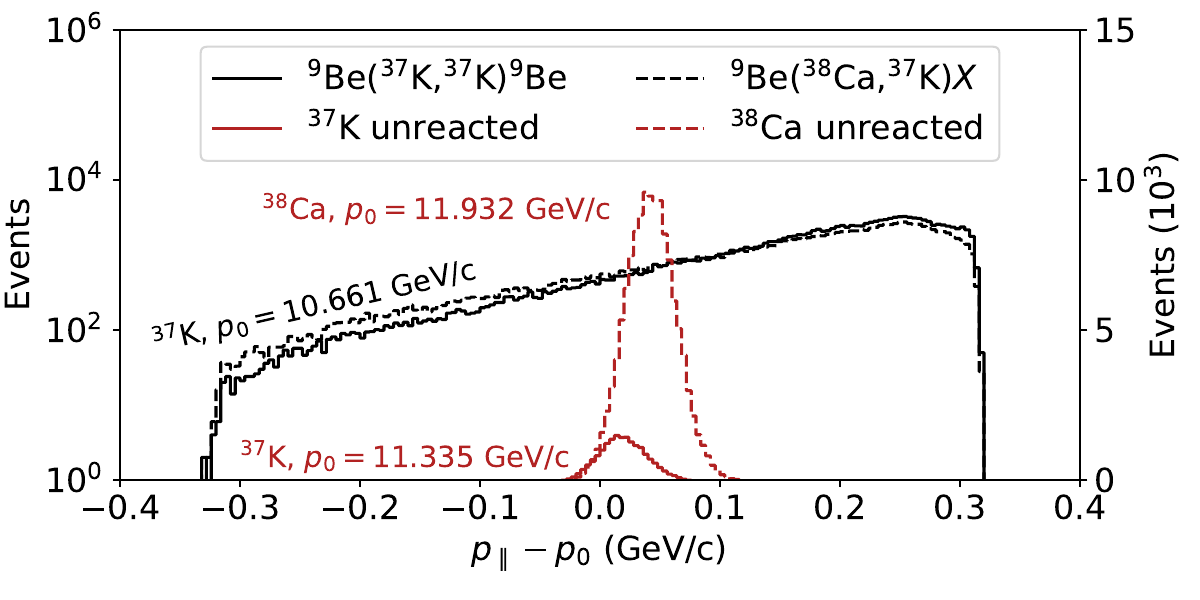}
\caption[]{
Measured parallel-momentum distributions of the \nuc{37}{K} residues from the 
\nuc{9}{Be}(\nuc{37}{K},$X$)$Y$ (solid black) and 
\nuc{9}{Be}(\nuc{38}{Ca},$X$)$Y$ (dashed black) reactions using the logarithmic scale on the left.
Their exponential character reveals the large momentum loss of roughly $675$\,MeV/c.
Parallel-momentum distributions for unreacted \nuc{37}{K} and \nuc{38}{Ca} ions, 
which suffered only target-related momentum losses, are shown in red for comparison
and use the vertical axis on the right side.
The given values for $p_0$ indicate the central momentum of the individual isotopes
in a given magnetic-rigidity setting of the S800 magnetic spectrograph.
}
\label{fig:dta}
\end{figure}

\begin{figure*}[t]
\centering
\includegraphics[trim=0 0 0 0,width=1.025\linewidth,clip]{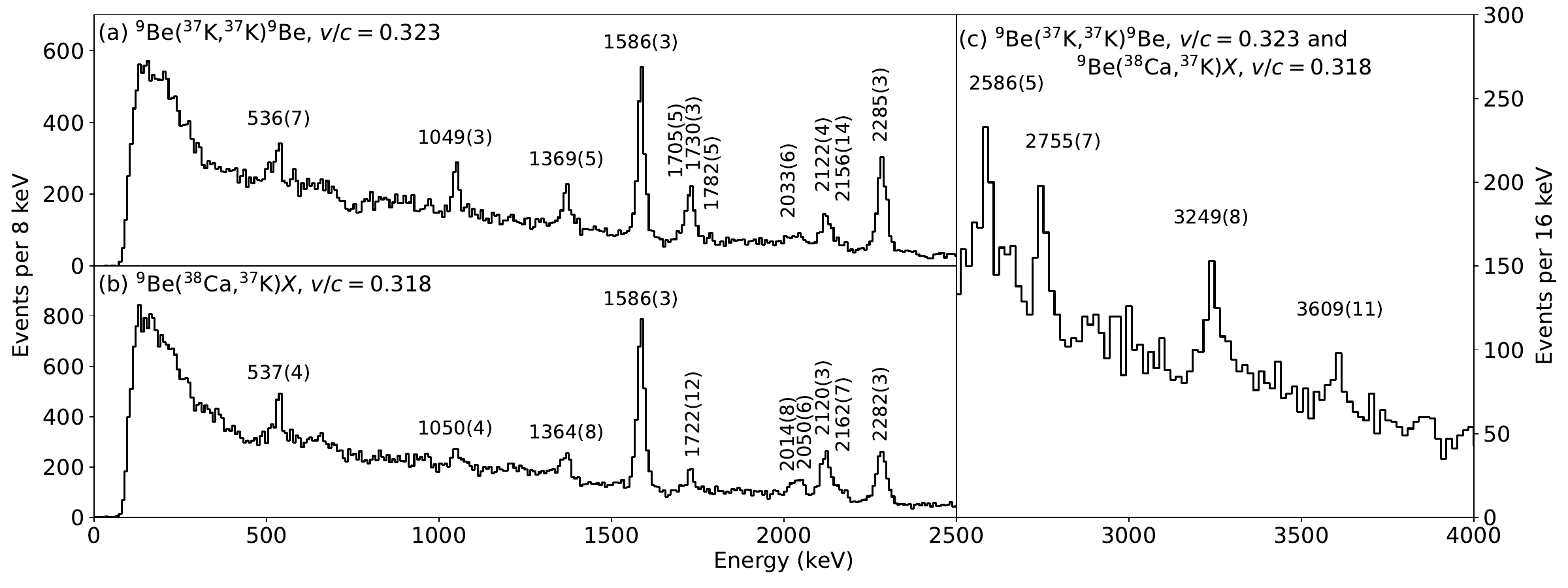}
\caption[]{
Doppler-corrected add-back $\gamma$-ray spectra in coincidence with \nuc{37}{K} residuals identified in the S800 focal plane 
from (a) the \nuc{9}{Be}(\nuc{37}{K},\nuc{37}{K}$+\gamma$)\nuc{9}{Be} and (b)
\nuc{9}{Be}(\nuc{38}{Ca},\nuc{37}{K}$+\gamma$)$X$ reactions.
In panel~(c), corresponding to energies above $2.5$\,MeV, the spectra of both reaction channels have been added together.
All identified $\gamma$-ray transitions are labeled by their energy.
}
\label{fig:spec}
\end{figure*}

With the given magnetic-rigidity setting, which was optimized for two-neutron removal from \nuc{38}{Ca} reported in Ref.~\cite{Bec23a},
only those \nuc{37}{K} nuclei undergoing a substantial momentum loss in the target reached the large-acceptance focal plane of the S800.
From Fig.~\ref{fig:dta}, it becomes obvious that they lie in the exponential low-momentum tail of the full parallel-momentum distribution;
in the case of the \mbox{\nuc{9}{Be}(\nuc{37}{K},\nuc{37}{K})\nuc{9}{Be}} reaction, roughly $675$\,MeV/c below the momenta of ions suffering 
target-induced energy loss only.
The high resolution $\gamma$-ray spectrometer GRETINA~\cite{Pas13a,Wei17a}, 
consisting of 12 detector modules with four 36-fold segmented high-purity germanium crystals each, 
was placed around the target for the detection of prompt $\gamma$ rays emitted in flight by the excited reaction products.
The interaction point with the highest energy deposition, as obtained from online pulse-shape analysis, 
was used for an event-by-event Doppler correction. 
Furthermore, information on the momentum vector of beam-like ions traced through the S800 spectrograph was included.
The resulting $\gamma$-ray spectra for both reactions are shown in Fig.~\ref{fig:spec}, 
using nearest-neighbor add-back introduced in Ref.~\cite{Wei17a}.

\begin{figure}[th]
\centering
\includegraphics[trim=0 0 0 0,width=1.025\linewidth,clip]{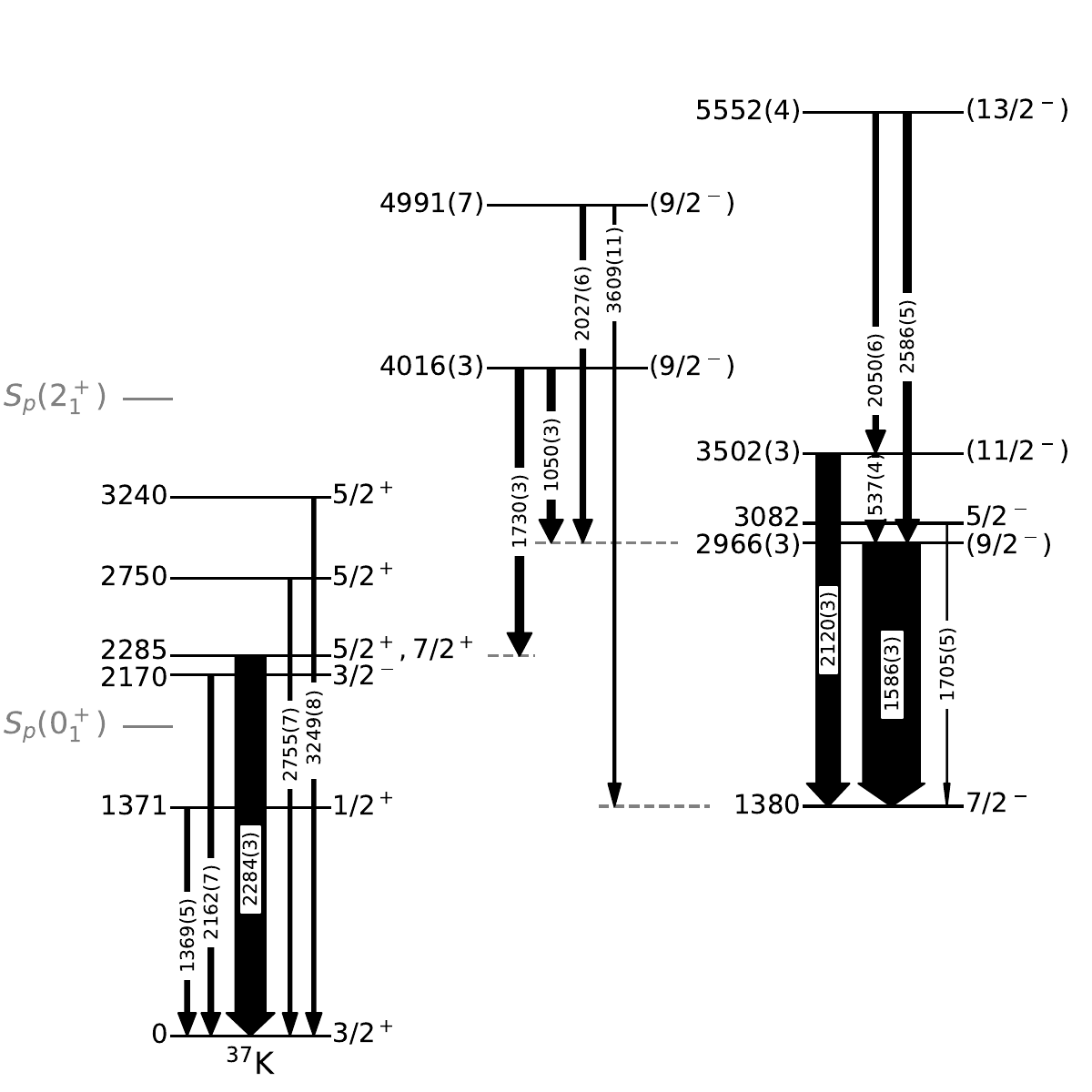}
\caption[]{Level scheme of \nuc{37}{K} as populated in the present work.
All excited states apart from the $1/2^+$ and $7/2^-$ states
at $1370.85(2)$ and $1380.25(3)$\,keV, respectively, 
are located above the proton-separation threshold to the ground state of \nuc{36}{Ar} $S_p(0^+_1)$, 
which is indicated by a grey line, and, thus, are proton unbound.
States above $S_p(2^+_1)$ can furthermore proton decay to the $2^+_1$ state of \nuc{36}{Ar}.
The arrow widths scale with the relative $\gamma$-ray intensities obtained 
from the \nuc{9}{Be}(\nuc{38}{Ca},\nuc{37}{K}$+\gamma$)$X$ reaction (cf. Tab.~\ref{tab:gamma_int}).
All energies are given in keV. Energies of excited states without uncertainties are taken from Ref.~\cite{NDS37}. 
Angular-momentum and parity quantum numbers in parentheses are tentative.
}
\label{fig:level}
\end{figure}

In the analysis,\footnote{In this work, uncertainties are propagated in the form of a Monte-Carlo method~\cite{GUM2} 
and might be asymmetrically distributed.
In the following, such probability distributions are characterized by their mode, i.e. the most probable value,
and the lower and upper boundaries of the shortest coverage interval.
The latter is constructed to contain 68.27\,\% of the distribution
and, thus, the given uncertainties are comparable to the $1\sigma$ interval of a normal distribution.} 
$\gamma$-ray energies below $2.5$\,MeV were determined individually for each reaction channel
with their weighted average used in the ensuing discussion.
Above $2.5$\,MeV, corresponding to panel~(c) of Fig.~\ref{fig:spec}, they are obtained from the summed spectrum.
Intensity ratios of $\gamma$-ray transitions given in the following are uncertainty-weighted averages over both reaction channels.
The spectra feature peaks at \onethreesevenzero, \twoonesevenzero, \twotwoeightfive, \twosevenfivezero, and \threetwothreenine\,keV
stemming from known transitions to the $3/2^+$ ground state of \nuc{37}{K}~\cite{NDS37}.
In coincidence with \nuc{37}{K} nuclei excited in the \nuc{9}{Be}(\nuc{37}{K},\nuc{37}{K}$+\gamma$)\nuc{9}{Be} reaction,
a depopulating transition of the $5/2^-$ state at $3081.99(9)$\,keV~\cite{NDS37} is observed at \onesevenzerotwo\,keV
whereas its remaining, weaker branching transitions cannot be resolved from the background.
The hitherto unknown $\gamma$ rays, including the most prominent peak at \onefiveeightseven\,keV,
are predominantly attributed to decays of proton-unbound -- $S_p(\nuc{37}{K})=1857.63(9)$\,keV~\cite{Wan21a} -- states 
into the isomeric $7/2^-$ state at $1380.25(3)$\,keV~\cite{NDS37}.
\begin{figure}[t]
\centering
\includegraphics[trim=0 0 0 0,width=1.025\linewidth,clip]{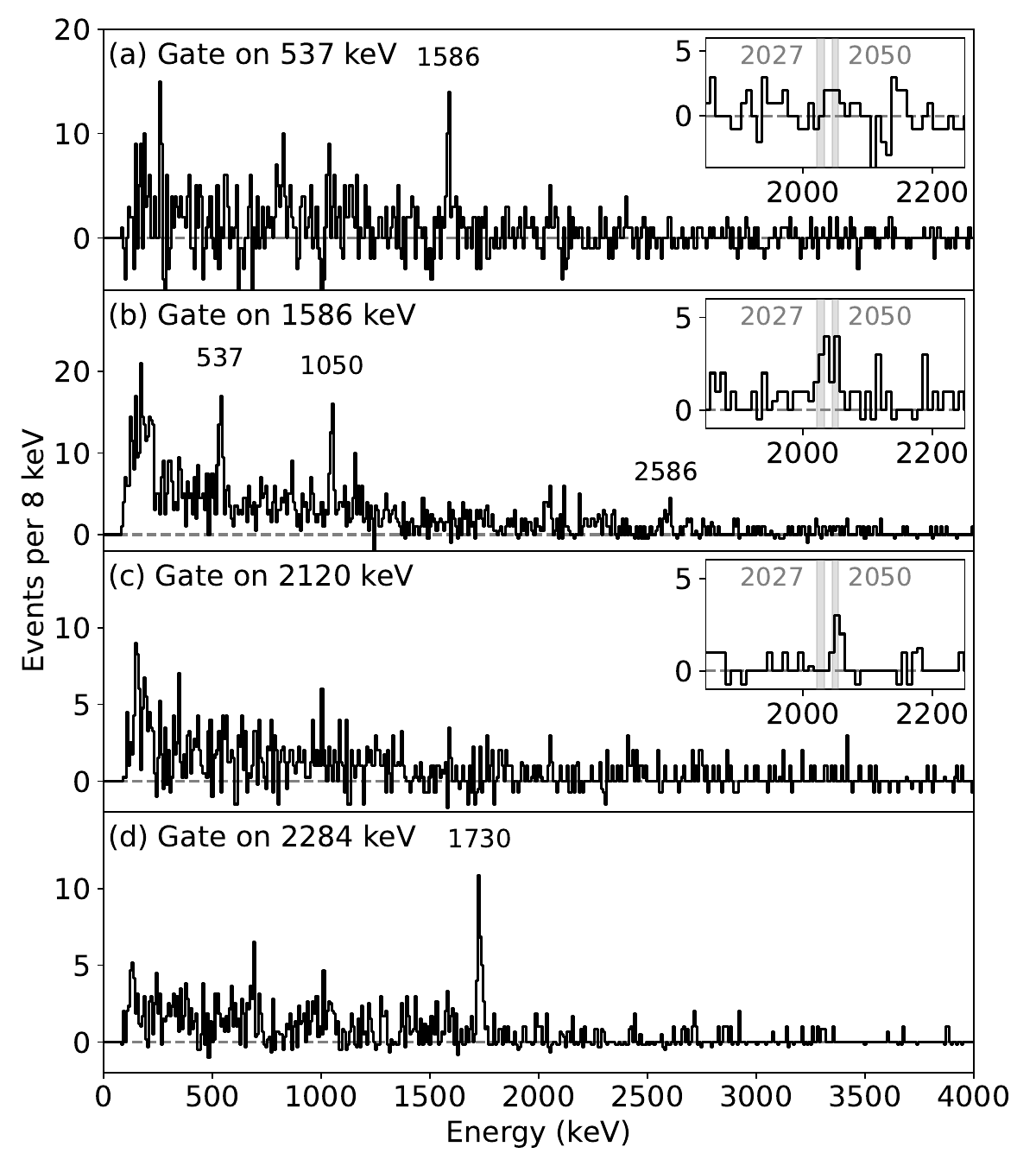}
\caption[]{Doppler-corrected, add-back $\gamma\gamma$ coincidence spectra 
from gates on prominent transitions in the spectra shown in Fig.~\ref{fig:spec}.
The coincidence relations deduced from the data displayed here, 
which are the sum of both reaction channels,
were used for the construction of a level scheme.
The insets enlarge the region of the \twozerotwofive- and \twozerofiveone-keV doublet 
and show data from the \nuc{9}{Be}(\nuc{38}{Ca},\nuc{37}{K}$+\gamma$)$X$ reaction.
They highlight that only the latter $\gamma$ ray is coincident with \fivethreesix~and \twoonetwoone\,keV.
}
\label{fig:coinc}
\end{figure}
Due to its long half life of $10.4(5)$\,ns~\cite{NDS37}, the isomer's depopulating $\gamma$ rays are emitted too far behind the target
for a meaningful Doppler reconstruction, preventing a direct observation in the spectra of Fig.~\ref{fig:spec}.
Employing the $\gamma\gamma$-coincidence relationships inferred from Fig.~\ref{fig:coinc}, 
all $\gamma$-ray transitions identified in the two reactions are placed in the level scheme presented in Fig.~\ref{fig:level},
with arrow widths scaled according to the intensities summarized in Tab.~\ref{tab:gamma_int}.
\begin{table}[tb]
\caption{Relative $\gamma$-ray intensities $I_{\gamma}$ of transitions with energy $E_{\gamma}$
from excited states of \nuc{37}{K} populated in the \nuc{9}{Be}(\nuc{37}{K},\nuc{37}{K}$+\gamma$)\nuc{9}{Be} and
\nuc{9}{Be}(\nuc{38}{Ca},\nuc{37}{K}$+\gamma$)$X$ reactions.
The corresponding $\gamma$-ray spectra are shown in Fig.~\ref{fig:spec}.
Transitions reported in Ref.~\cite{NDS37} are marked by an asterisk.}
\label{tab:gamma_int}
\begin{ruledtabular}
\begin{tabular}{rcc}
$E_{\gamma}$ (keV)			&\multicolumn{2}{c}{$I_{\gamma}$ ($\%$)}\\
						&\nuc{9}{Be}(\nuc{37}{K},\nuc{37}{K})\nuc{9}{Be}	&\nuc{9}{Be}(\nuc{38}{Ca},\nuc{37}{K})$X$\\\colrule
\fivethreesix\phantom{$^*$}			&$6(3)$									&$9(2)$\\		
\onezerofournine\phantom{$^*$}		&$22(3)$									&$13(2)$\\
\onethreesevenzero$^*$				&$15_{-3}^{+4}$							&$7(3)$\\
\onefiveeightseven\phantom{$^*$}		&$100$									&$100$\\
\onesevenzerotwo$^*$				&$7_{-3}^{+2}$								&$1(1)$\\
\oneseventhreezero\phantom{$^*$}		&$38_{-4}^{+5}$							&$14(3)$\\
\oneseveneightfour\phantom{$^*$}		&$7(2)$									&$$\\
\twozerotwofive\phantom{$^*$}			&$7_{-3}^{+4}$								&$8(3)$\\
\twozerofiveone\phantom{$^*$}			&$1_{-1}^{+3}$								&$8(3)$\\
\twoonetwoone\phantom{$^*$}			&$30(4)$									&$41(4)$\\
\twoonesevenzero$^*$				&$5(3)$									&$8(2)$\\
\twotwoeightfive$^*$					&$82_{-6}^{+7}$							&$53(4)$\\
\twofiveeightseven\phantom{$^*$}		&$2(2)$									&$12(2)$\\
\twosevenfivezero$^*$				&$12(3)$									&$5(2)$\\
\threetwothreenine$^*$				&$9(3)$									&$6(2)$\\
\threesixonetwo\phantom{$^*$}			&$6(2)$									&$4(2)$\\
\end{tabular}
\end{ruledtabular}
\end{table}
The excited states reported here for the first time will be discussed in the following.

The most intense $\gamma$ ray is placed as the transition from a state at \twoninesixsix\,keV to the $7/2^-$ isomeric state.
The former is only known from an experiment employing the \nuc{40}{Ca}$(p,\alpha)$ reaction~\cite{Mag95a}
and is identified as the mirror of the $9/2^-$ state of \nuc{37}{Ar} at $3184.74(19)$\,keV~\cite{NDS37}.
Utilizing $\gamma\gamma$ coincidences, the $\gamma$-ray transitions at 
\fivethreesix, \onezerofournine, \twozerotwofive, and \twofiveeightseven\,keV are found to feed this state
[cf. panel~(b) of Fig.~\ref{fig:coinc}].
The first depopulates a hitherto unknown state at \threefivezerotwo\,keV which is tentatively assigned
quantum numbers \hbox{$J^{\pi}=11/2^-$}, assuming mirror symmetry with the $3706.19(23)$-keV state of \nuc{37}{Ar}~\cite{NDS37}.
It furthermore features an intense \twoonetwoone-keV transition to the $7/2^-$ state at $1380.25(3)$\,keV.
The intensity ratio $I_{\gamma}(537)/I_{\gamma}(2120)=0.23(5)$
is in agreement with $0.177(12)$ found for the corresponding transitions in \nuc{37}{Ar}~\cite{NDS37}.

The $\gamma$ ray observed at \onezerofournine\,keV is 
only coincident with the \onefiveeightseven-keV transition.
Thus, it is proposed to originate from an excited state of \nuc{37}{K} at \fourzeroonethree\,keV.
It might coincide with a state at $4018(5)$\,keV observed in the \nuc{40}{Ca}$(p,\alpha)$ reaction 
for which a dominant electromagnetic decay channel is reported~\cite{Mag95a}.
An additional depopulating transition to the $(5/2^+,7/2^+)$ state at $2285.24(12)$\,keV~\cite{NDS37},
corresponding to a $\gamma$-ray energy of \oneseventhreezero\,keV, and 
an intensity ratio \mbox{$I_{\gamma}(1050)/I_{\gamma}(1730)=0.61(10)$}
singles it out as the potential mirror for the $9/2^-$ state of \nuc{37}{Ar} located at $4021.6(3)$\,keV~\cite{NDS37}.
The latter also features a weak decay branch to the isomeric $7/2^-$ state 
which is not resolved from the background in the present study of \nuc{37}{K} (cf. Fig.~\ref{fig:spec}).

From the coincidences between the \onefiveeightseven- and \twofiveeightseven-keV $\gamma$ rays 
the existence of an excited state at \fivefivefivethree\,keV is concluded. 
As shown in Fig.~\ref{fig:coinc} it also features a branching transition to the $(11/2^-)$ state at \threefivezerotwo\,keV
with a $\gamma$-ray intensity ratio $I_{\gamma}(2050)/I_{\gamma}(2586)=0.63_{-0.22}^{+0.28}$.
The only known excited state of \nuc{37}{Ar} with these decay characteristics is the $13/2^-$ state at $5793.3(3)$\,keV~\cite{NDS37}.
The branching ratio  of its decay transitions to the $11/2^-$ and $9/2^-$ states, 
which has recently been corrected to $0.67(4)$~\cite{Das20a}, is in excellent agreement with the results for \nuc{37}{K}.

For the \fournineninetwo-keV state of \nuc{37}{K} $\gamma$-ray transitions into the 
$7/2^-$ and $(9/2^-)$ states at $1380.25(3)$ and \twoninesixsix\,keV are observed, respectively.
From the literature data available for \nuc{37}{Ar}~\cite{NDS37,Das20a}, no state with matching decay pattern is known.
They either feature dominant decay branches not observed here or, 
in the case of a $7/2^-,11/2^-$ state at $4981.0(6)$\,keV~\cite{NDS37} 
only populated in the \nuc{34}{S}$(\alpha,n\gamma)$~\cite{Nol75a} and \nuc{35}{Cl}(\nuc{3}{He}$,p$)~\cite{Mcn66a} reactions,
lack the $\gamma$-ray transition to the $9/2^-$ state at $3184.74(19)$\,keV~\cite{Nol75a}.
Its interpretation needs additional input from theory which is presented in the following Section.

Finally, a weak $\gamma$-ray transition at \oneseveneightfour\,keV is observed in the $\gamma$-ray spectra
following inelastic scattering of \nuc{37}{K} ions [cf. panel~(a) of Fig.~\ref{fig:spec}].
No information on it is available in the data sheets and a placement in the level scheme from a coincidence
analysis akin to Fig.~\ref{fig:coinc} is not possible due to low statistics.
Thus, it remains unplaced in the level scheme shown in Fig.~\ref{fig:level}.

While a comparison of excitation energies of the \nuc{37}{K}/\nuc{37}{Ar} mirror pair is already found in Ref.~\cite{Esc88a},
the extension of the level scheme from the present work warrants a new review of mirror-energy differences reported in Tab.~\ref{tab:med}.

\begin{table*}[t]
\caption{Compilation of excitation energies $E_x$ and mirror-energy differences for states 
of the mirror pair \nuc{37}{K}/\nuc{37}{Ar} observed in this work.
The levels are grouped in the same way as in Fig.~\ref{fig:level}. 
Excitation energies are compared to shell-model predictions $E_x^{\text{th}}$ 
using the FSU effective interaction~\cite{Lub19a,Lub20a},
which furthermore yield occupation numbers for \nuc{37}{K} 
of the $1d_{3/2}$, $2s_{1/2}$, $1f_{7/2}$, and $2p_{3/2}$ orbitals.
Due to the isospin-invariant nature of the interaction the occupancies of \nuc{37}{Ar} 
can be obtained through exchange of the proton and neutron labels.
}
\label{tab:med}
\begin{ruledtabular}
\begin{tabular}{clllccc}
$J^{\pi}$				&\multicolumn{2}{c}{$E_x$ (keV)}			&MED (keV)				&$E_x^{\text{th}}$ (keV)	&\multicolumn{2}{c}{Occupation numbers \nuc{37}{K}}\\
					&\nuc{37}{K}			&\nuc{37}{Ar}		&						&					&$\nu(d_{3/2},~s_{1/2},~f_{7/2},~p_{3/2})$					&$\pi(d_{3/2},~s_{1/2},~f_{7/2},~p_{3/2})$\phantom{$^2$}\\\colrule
$3/2^+$\phantom{$^1$}	&$\phantom{000}0$		&$\phantom{000}0$	&$\phantom{-000}0$			&\phantom{000}$0$		&$\phantom{\nu}(2.21,~1.90,~0.00,~0.00)$		&$\phantom{\pi}(3.07,~1.96,~0.00,~0.00)$\phantom{$^2$}\\
$1/2^+$\phantom{$^1$}	&$1370.85(2)$			&$1409.84(7)$		&$-\phantom{3}38.99(7)$		&1389				&$\phantom{\nu}(2.60,~1.52,~0.00,~0.00)$		&$\phantom{\pi}(3.53,~1.56,~0.00,~0.00)$\phantom{$^2$}\\
$3/2^-$\phantom{$^1$}	&$2170.18(13)$		&$2490.17(13)$	&$-320.00(20)$				&2650				&$\phantom{\nu}(2.43,~1.78,~0.01,~0.00)$		&$\phantom{\pi}(2.48,~1.78,~0.26,~0.70)$\phantom{$^2$}\\
$7/2^+$$^1$			&$2285.24(12)$		&$2217.00(23)$	&$+\phantom{3}68.24(26)$	&2135				&$\phantom{\nu}(2.12,~1.93,~0.00,~0.00)$		&$\phantom{\pi}(3.02,~1.99,~0.00,~0.00)$\phantom{$^2$}\\
$5/2^+$\phantom{$^1$}	&$2750.22(8)$			&$2796.15(8)$		&$-\phantom{3}45.93(12)$	&2770				&$\phantom{\nu}(2.18,~1.92,~0.00,~0.00)$		&$\phantom{\pi}(3.16,~1.93,~0.00,~0.00)$\phantom{$^2$}\\
$5/2^+$\phantom{$^1$}	&$3239.5(2)$			&$3170.0(5)$		&$+\phantom{3}69.5(6)$		&3154				&$\phantom{\nu}(2.59,~1.58,~0.00,~0.00)$		&$\phantom{\pi}(3.47,~1.72,~0.00,~0.00)$\phantom{$^2$}\\\colrule
$9/2^-$\phantom{$^1$}	&\fourzeroonethree		&$4021.6(3)$		&$-\phantom{33}6(3)$		&4093				&$\phantom{\nu}(2.07,~1.68,~0.39,~0.01)$		&$\phantom{\pi}(2.90,~1.68,~0.46,~0.01)$$^2$\\
$9/2^-$\phantom{$^1$}	&\fournineninetwo		&				&$$						&5240				&$\phantom{\nu}(2.11,~1.67,~0.38,~0.00)$		&$\phantom{\pi}(2.86,~1.69,~0.54,~0.02)$\phantom{$^2$}\\\colrule
$7/2^-$\phantom{$^1$}	&$1380.25(3)$			&$1611.28(5)$		&$-231.03(6)$				&1545				&$\phantom{\nu}(2.57,~1.67,~0.02,~0.00)$		&$\phantom{\pi}(2.58,~1.69,~0.96,~0.01)$\phantom{$^2$}\\
$9/2^-$\phantom{$^1$}	&\twoninesixsix			&$3184.74(19)$	&$-219(3)$				&2913				&$\phantom{\nu}(2.59,~1.60,~0.06,~0.00)$		&$\phantom{\pi}(2.61,~1.67,~0.92,~0.00)$$^3$\\
$5/2^-$\phantom{$^1$}	&$3081.99(9)$			&$3273.58(14)$	&$-191.59(17)$				&2844				&$\phantom{\nu}(2.63,~1.54,~0.08,~0.01)$		&$\phantom{\pi}(2.85,~1.47,~0.84,~0.02)$\phantom{$^2$}\\
$11/2^-$\phantom{$^1$}	&\threefivezerotwo		&$3706.19(23)$	&$-204(3)$				&3761				&$\phantom{\nu}(2.45,~1.70,~0.09,~0.00)$		&$\phantom{\pi}(2.55,~1.75,~0.87,~0.01)$\phantom{$^2$}\\
$13/2^-$\phantom{$^1$}	&\fivefivefivethree		&$5793.3(3)$		&$-241(4)$				&5786				&$\phantom{\nu}(2.67,~1.64,~0.07,~0.00)$		&$\phantom{\pi}(2.68,~1.74,~0.90,~0.00)$\phantom{$^2$}\\
\end{tabular}
\end{ruledtabular}
$^1$ For the $2285.24(12)$-keV state of \nuc{37}{K} $J^{\pi}=5/2^+,7/2^+$ is found in Ref.~\cite{NDS37}.\\
$^2$ The $9/2^-_2$ state has the structure of $1d_{3/2}$ coupled to the $3^-_1$ state of \nuc{36}{Ar}.\\
$^3$ The $9/2^-_1$ state has the structure of $1f_{7/2}$ coupled to the $2^+_1$ state of \nuc{36}{Ar}. 
\end{table*}

\section{Discussion}

\begin{figure}[t]
\centering
\includegraphics[trim=325 50 325 50,width=1.025\linewidth,clip]{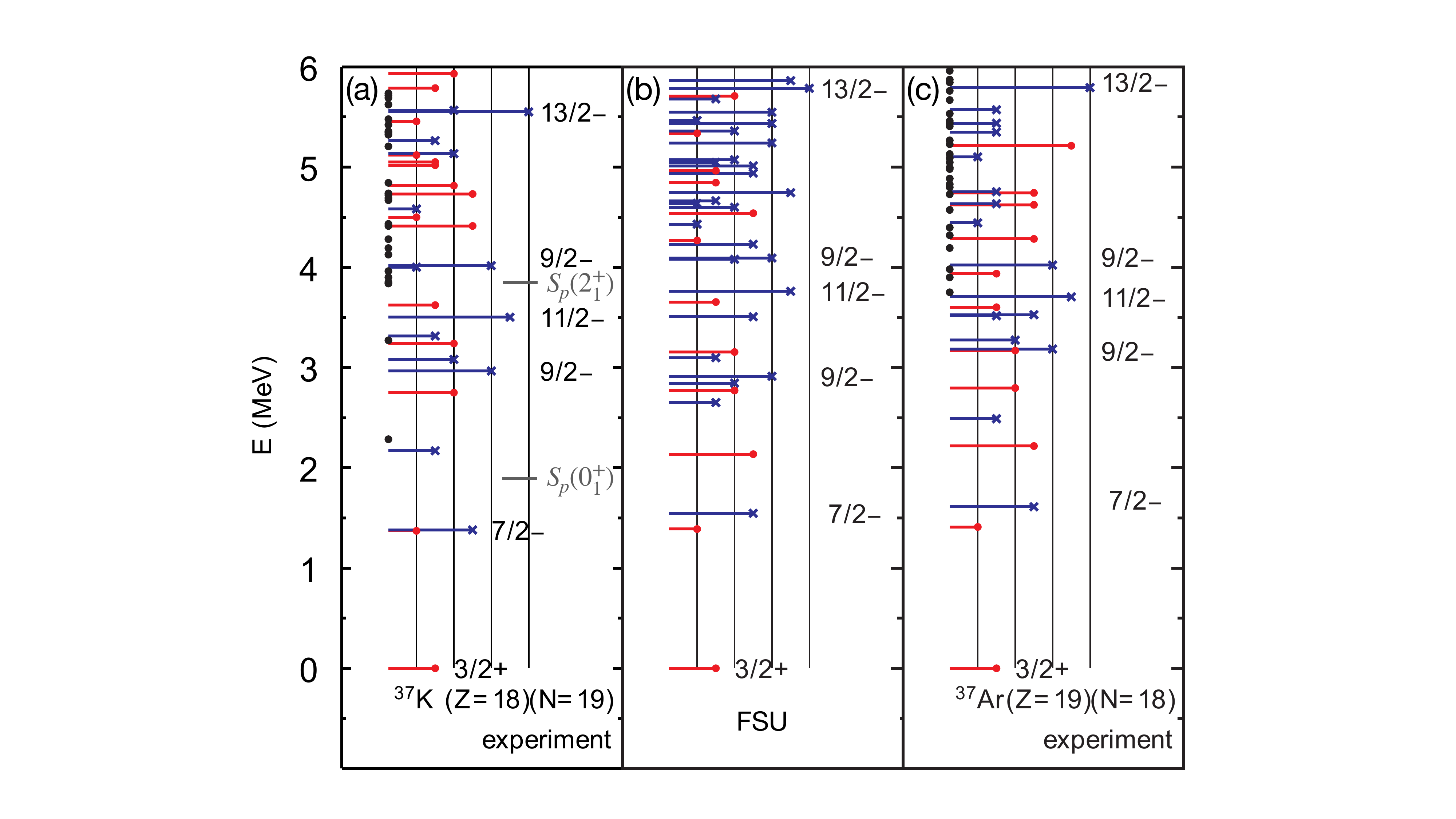}
\caption[]{Excitation energies of low-lying states of the mirror nuclei (a) \nuc{37}{K} and (c) \nuc{37}{Ar}
compared to (b) shell-model calculations using the FSU cross-shell interaction.
The angular-momentum quantum number is revealed by the length of the line
and positive (negative) parity is indicated by red (blue) color.
Black dots in the panels displaying experimental data denote states with unknown quantum numbers.
The proton-separation thresholds of \nuc{37}{K} to the $0^+_1$ and $2^+_1$ states of \nuc{36}{Ar} 
are indicated by grey lines.}
\label{fig:fsu}
\end{figure}

The assignment of angular momentum and parity quantum numbers $J^{\pi}=9/2^-$, $11/2^-$, and $13/2^-$
based on isospin symmetry between \nuc{37}{K} and \nuc{37}{Ar} is supported by the conclusions of Refs.~\cite{Gad22a,Gad22b};
stating that higher-spin, negative-parity, complex-structure states are populated in 
fast-beam induced, highly-momentum dissipative processes exploited in the present work.
These states lie, due to their unnatural parity, outside of the model space of $sd$-shell limited Hamiltonians 
such as USDB~\cite{Bro06a} or USDC~\cite{Mag20a}.
Instead, their shell-model description necessitates the inclusion of cross-shell excitations either from the $1p$ or,
in the present case of \nuc{37}{K}, more importantly into the $1f$-$2p$ major shells.
The FSU interaction~\cite{Lub19a,Lub20a}, which employs the extensive $spsdf\!p$ model space,
is a recent representative of the class of cross-shell interactions.
Its comparison for $A=37$ to the level schemes of \nuc{37}{K} and \nuc{37}{Ar} is presented in Fig.~\ref{fig:fsu}.
Similar to the neighboring, heavier $A=38$~\cite{Gad22b} and $39$~\cite{Gad22a} cases, the agreement is good.\\

From the occupation numbers obtained in the calculations, which are collected in Tab.~\ref{tab:med},
the negative-parity states above the isomeric $7/2^-$ state at $1380.25(3)$\,keV (cf. Fig.~\ref{fig:level})
are revealed as almost pure $1p-1h$ states with one proton promoted into the $1f_{7/2}$ orbit.
Their mirror-energy differences $\Delta E=E^{N<Z}-E^{N>Z}$ exceed $-190$\,keV
which places them within the range normally attributed to Coulomb-interaction induced isospin breaking~\cite{Hen20a}
if the larger spatial extent of the wave functions due to proton-unbound nature of these states in \nuc{37}{K} is considered.
The relative MEDs principally originate in the different sizes of the valence orbitals and 
relativistic electromagnetic spin-orbit corrections~\cite{Nol69a}.
The MEDs for the $7/2^-$ and $3/2^-$ states are negative, 
qualitatively because the root mean square (rms) radii of the $fp$-shell valence orbitals are larger than
the $sd$-shell valence orbitals, and the Coulomb valence-core interaction becomes relatively smaller.
In a finite potential well the rms radius of the $2p_{3/2}$ orbital is larger than
that for the $1f_{7/2}$ orbital because of the smaller angular-momentum barrier for $l=1$ compared to $l=3$. 
Thus, the valence-core Coulomb interaction is smaller for $2p_{3/2}$ than $1f_{7/2}$, 
resulting in a Thomas-Ehrman shift for low-$l$ orbitals~\cite{Ehr51a,Tho52a}.
This is observed in Tab.~\ref{tab:med} for the $3/2^-$ state.
It qualitatively explains the observed MEDs for the \nuc{37}{K}/\nuc{37}{Ar} mirror pair.
Furthermore, it has to be noted that the structure of the second and third $9/2^-$ levels 
is different from all other states in Tab.~\ref{tab:med} 
in that they have about an equal fraction of excited $1f_{7/2}$ protons and neutrons 
resulting in a displacement energy that is similar to the \nuc{37}{K}/\nuc{37}{Ar} ground state 
and an MED of only $-6(3)$\,keV for the $9/2^-_{2}$ state.
More quantitative calculations that take into account the many-body Coulomb~\cite{Zuk02a,Duf02a} 
and electromagnetic spin-orbit corrections are beyond the scope of this paper.

The value of the proton separation thresholds for \nuc{37}{K} to the $0^+_1$ and $2^+_1$ states of \nuc{36}{Ar},
$S_p(0^+_1)=1857.63(9)$~\cite{Wan21a} and $S_p(2^+_1)=3828.01(10)$\,keV~\cite{Wan21a,NDS36}, respectively,
are shown in panel~(a) of Fig.~\ref{fig:fsu}. 
An evaluation of the $\gamma$ decays observed for states above $S_p(0^+_1)$ and $S_p(2^+_1)$ must thus consider estimates 
of the partial proton decay half-lives to the $0^+$ ground and first excited $2^+_1$ states of \nuc{36}{Ar}, respectively.
The calculated $\gamma$-ray emission and proton-decay properties can be used to shed some light on the
newly found excited state of \nuc{37}{K} at \fournineninetwo\,keV.
The shell-model results place $3/2^-$ and $5/2^-$ states at $5048$ and $5073$\,keV, respectively. 
For both the $\gamma$-decay pattern is not consistent with the experimental observation (cf. Fig.~\ref{fig:level}).
Two $7/2^-$ states in the vicinity ($4939$ and $5012$\,keV) are predicted to predominantly proton decay 
to the $2^+_1$ and $0^+_1$ states of \nuc{36}{Ar} with $l=1$ and $3$ and 
spectroscopic factors $C^2S=0.024$ and $0.015$, respectively.
This is supported by experimentally determined $\gamma$-to-proton branching ratios which are reported to vanish
for excited states in this energy region with angular momentum quantum numbers up to $J=7/2$~\cite{Mag95a}.
The $9/2^-_3$ and $11/2^-_2$ states with calculated excitation energies of $5240$ and $4744$\,keV 
are the only other $1p$-$1h$ states in the region of interest.
The latter is most likely not observed in the present experiment since it decays under emission of a proton 
from the $f_{7/2}$ orbital with $C^2S=0.035$.
In contrast, the spectroscopic factor of this reaction for the $9/2^-_3$ state is found to be $0.008$,
making the existence of a sizeable $\gamma$-decay branch reasonable.
Its predicted electromagnetic decay behavior features dominant transitions to the $7/2^-_1$
and $9/2^-_1$ states with an intensity ratio of roughly $0.5$.
The corresponding transitions are indeed observed in the experimental data (cf. Fig.~\ref{fig:level}),
however, with the intensity ratio not well constrained due to low statistics.
Hence, an assignment $J^{\pi}=9/2^-$ is tentatively proposed for the state at \fournineninetwo\,keV.
Based on the composition of its wave function given in Tab.~\ref{tab:med}, 
which is predicted to be similar to the $(9/2^-)$ state at \fourzeroonethree\,keV,
only a small energy displacement compared to \nuc{37}{Ar} is expected.
As already mentioned in Section~\ref{sec:res}, no suitable state, 
i.e. with the right excitation energy and quantum numbers, 
is found in the level scheme of \nuc{37}{Ar}~\cite{NDS37,Das20a}.
New experimental data on the $4981.0(6)$-keV state, 
which was not populated in the \nuc{27}{Al}(\nuc{12}{C},$np$)\nuc{37}{Ar} reaction used in Ref.~\cite{Das20a},
either in the form of an unambiguous determination of its angular-momentum and parity quantum numbers
or an identification of its potential $\gamma$-decay branch to the $9/2^-$ state at $3184.74(19)$\,keV,
might clarify the situation.

\section{Summary}

In summary, excited states of the neutron-deficient nucleus \nuc{37}{K} were populated 
in inelastic-scattering and proton-removal reactions at high momentum loss.
From subsequent in-beam $\gamma$-ray spectroscopy, an extension of the level scheme 
up to the yrast $(13/2^-)$ state is proposed.
Shell-model calculations using the FSU cross-shell effective interaction describe the structure
of the \nuc{37}{K}/\nuc{37}{Ar} mirror pair remarkably well and highlight the importance of cross-shell excitations 
into the $pf$ shell for the interpretation of excited states of \nuc{37}{K}.
The obtained occupation numbers of $sd$- and $pf$-shell orbitals suggest an almost pure proton $1p-1h$ nature 
of the states on top of the isomeric $7/2^-$ state of \nuc{37}{K}.
Two $(9/2^-)$ states with comparable occupancies in the proton and neutron $1f_{7/2}$ orbitals
are identified above $4$\,MeV; the former of which exhibits only a small mirror-energy difference.
The observed population pattern, which favors higher-spin, negative-parity states, 
adds dissipative reactions induced by intermediate energy rare-isotope beams
to the experimental toolbox for future studies of mirror-energy differences for high-spin and complex-structure states;
especially if traditional population pathways of such states through fusion-evaporation reactions are not viable.

\begin{acknowledgments}

This work was supported by the U.S. NSF under Grants No. PHY-1565546 and No. PHY-2110365, 
by the DOE NNSA through the NSSC, under Award No. DE-NA0003180, 
and by the U.S. DOE, Office of Science, Office of Nuclear Physics, under Grants No. DE-SC0020451 \& DE-SC0023633 (MSU) and No. DE-FG02-87ER-40316 (WashU) 
and under Contract No. DE-AC02-06CH11357 (ANL).
GRETINA was funded by the DOE, Office of Science. 
Operation of the array at NSCL was supported by the DOE under Grant No. DE- SC0019034.

\end{acknowledgments}

\bibliography{bib_37K}

\end{document}